\documentclass[pra,reprint,amsmath,amssymb,aps,floatfix]{revtex4-1}
\usepackage{dcolumn,graphicx,color}
\usepackage{upgreek}
\usepackage{pdfpages}

\def\ket#1{\mathinner{|{#1}\rangle}}

\begin{document}

\title{Towards High-Fidelity Quantum Computation and Simulation on a Programmable Photonic Integrated Circuit}

\author{Jacob Mower$^{*}$}
\affiliation{Department of Electrical Engineering and Computer Science, Massachusetts Institute of Technology, 77 Massachusetts Avenue, Cambridge, MA 02139, USA}

\author{Nicholas C. Harris$^{*}$}
\affiliation{Department of Electrical Engineering and Computer Science, Massachusetts Institute of Technology, 77 Massachusetts Avenue, Cambridge, MA 02139, USA}

\author{Gregory R. Steinbrecher$^{*}$}
\affiliation{Department of Electrical Engineering and Computer Science, Massachusetts Institute of Technology, 77 Massachusetts Avenue, Cambridge, MA 02139, USA}

\author{Yoav Lahini}
\affiliation{Department of Physics, Massachusetts Institute of Technology, 77 Massachusetts Avenue, Cambridge, MA 02139, USA}

\author{Dirk Englund}
\affiliation{Department of Electrical Engineering and Computer Science, Massachusetts Institute of Technology, 77 Massachusetts Avenue, Cambridge, MA 02139, USA}

\begin{abstract}
We propose and analyze the design of a programmable photonic integrated circuit for high-fidelity quantum computation and simulation. We demonstrate that the reconfigurability of our design allows us to overcome two major impediments to quantum optics on a chip: it removes the need for a full fabrication cycle for each experiment and allows for compensation of fabrication errors using numerical optimization techniques. Under a pessimistic fabrication model for the silicon-on-insulator process, we demonstrate a dramatic fidelity improvement for the linear optics CNOT and CPHASE gates and, showing the scalability of this approach, the iterative phase estimation algorithm built from individually optimized gates. We also propose and simulate a novel experiment that the programmability of our system would enable: a statistically robust study of the evolution of entangled photons in disordered quantum walks. Overall, our results suggest that existing fabrication processes are sufficient to build a quantum photonic processor capable of high fidelity operation.
\end{abstract}

\maketitle

\section{Introduction}
Photonic integrated circuits (PICs) --- waveguide-based systems of optical elements such as beamsplitters and phase shifters that are monolithically integrated on a single chip --- enable control over the propagation and coupling of optical modes with exceptional phase stability and at the scale of tens to hundreds of waveguides. In particular, PICs fabricated using mature silicon processes have seen rapid development in recent years for optical interconnects and other classical applications \cite{2010.vlasov.exascale,2000.miller.optical_intercon}.
Additionally, PICs have been shown to be an appealing platform for quantum optics: PIC-based experiments have demonstrated quantum simulation \cite{2012.NPhys.Walther.photonic_quantum_sim,2013.Crespi.Sciarrino.boson, 2010.Peruzzo.OBrien.QRW}, boson sampling \cite{2011.ACM.Aaronson.boson_sampling,2013.Broome.White.BosSamp,2013.Spring.Walmsley.BosSamp}, linear optical quantum gates \cite{2008.Science.OBrien.quantum_circuit,2012.NPhoton.Obrien.Shor_photon_recycling}, and the simulation of bosonic quantum walks \cite{PhysRevLett.100.013906,2010.Peruzzo.OBrien.QRW,ander.PIC.2013}.

One of the main impediments to quantum optics experiments on PICs has been the need to fabricate custom chips for each experiment, an expensive and time consuming process. In addition, many applications require PICs to be tuned between consecutive experiments. While some experiments have shown on-chip reconfigurability \cite{Matthews:2009uo, 2014.peruzzo.obrien.eigenvalue, 2012.shadbolt.obrien.source}, there has been to date no analysis of a fully reconfigurable PIC that can implement arbitrary circuits. Additionally, PIC-based experiments to date have suffered from reduced fidelity due to variations and imperfections in the fabrication process.

In this work, we propose and analyze the design of a reconfigurable quantum photonic processor (QPP) --- achievable with existing, mature silicon processes --- that overcomes fabrication imperfections. We demonstrate how to program arbitrary transformations into this system and, using a fabrication model with conservative assumptions on technology, demonstrate a tuning algorithm that overcomes fabrication imperfections and achieves high fidelity quantum operations. This programmable linear optics circuit would enable the rapid testing of quantum optics algorithms.

In the next section, we introduce the QPP architecture and discuss the origins of imperfections in realistic devices. Section III shows how to implement quantum gates on a QPP, quantifies the detrimental effects of fabrication errors, and then demonstrates a computationally scalable, gate-by-gate procedure that allows us to recover high-fidelity gate operation. As an example of the power of this technique, we analyze a circuit implementing a full quantum algorithm, the iterative phase estimation algorithm (IPEA), and show that gate-by-gate optimization is sufficient for high-fidelity operation of the full circuit. Next, in Section V, we propose and simulate a novel bosonic transport experiment that leverages the reconfigurability of the QPP to investigate 1000 realizations of quantum walks under a range of disorder and decoherence levels and to perform state preparation on a pair of input photons. To close, we discuss methods of extending this architecture with recent advances in integrated quantum devices.

\begin{figure*}[tb]
\includegraphics[width=6.5in]{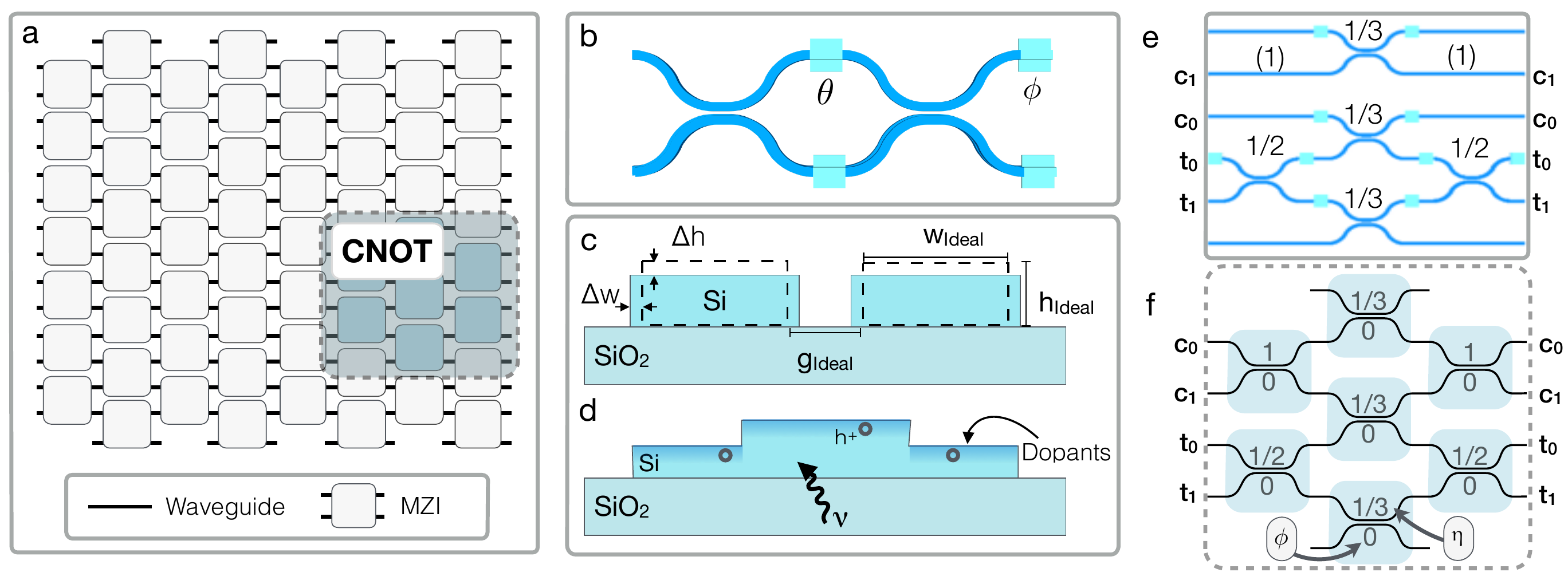}
\caption{{\footnotesize
(a) Schematic of the QPP composed of interconnected MZIs.
(b) The six-mode CNOT gate proposed in Ref. \cite{PhysRevA.65.062324}.
(c) The same CNOT protocol implemented on the QPP. The upper number in each box represents the splitting ratio $\eta \equiv \sin^{2}(\theta)$, where $\theta$ is the internal phase setting, and the lower number represents the output phase offset $\phi$.
(d) The MZI unit cell.
(e) Cross-section of the directional coupler showing the dominant mechanisms for disorder in the splitting ratio, variation in the height of the waveguide $h$, the width $w$, and the waveguide spacing $g$.
(f) Cross section of the phase shifter illustrating free carrier absorption, the dominant loss mechanism \cite{2014.opex.harris-galland.heater}.}}
\label{fig:QPP}
\end{figure*}

\section{The QPP Architecture}
\label{sec:hardware}

The proposed QPP architecture consists of a lattice of $2 \times 2$ building blocks (Figure \ref{fig:QPP}(a)), each of which is a Mach-Zehnder interferometer (MZI) (Figure \ref{fig:QPP}(b)) \cite{1994.PRL.Reck-Zeilinger-Bartani.any_unitary_operator}. In the spatial mode basis, an ideal MZI applies the $2 \times 2$ unitary given by
\begin{equation*}
U_{\text{MZI}}(\theta, \phi)=
\frac{1}{2}
\left(
\begin{matrix}
e^{i \phi} & 0\\
0 & 1
\end{matrix}
\right)
\left(
\begin{matrix}
1 & i\\
i & 1
\end{matrix}
\right)
\left(
\begin{matrix}
e^{i \theta} & 0\\
0 & 1
\end{matrix}
\right)
\left(
\begin{matrix}
1 & i\\
i & 1
\end{matrix}
\right),
\end{equation*}
where $\theta$ and $\phi$ correspond to the labels in Figure \ref{fig:QPP}(b).

In realistic integrated optical devices, photon loss, phase errors, and unbalanced beam splitters can severely impact performance.
To simulate the effect of these imperfections, we consider a model for the well developed, CMOS-compatible silicon-on-insulator (SOI) platform, based on deep-UV photolithography \cite{2012.Baehr-Jones.Hochberg.25GBS, Hochberg:2013vj}.
As photon loss is a primary concern in quantum optics experiments, we have chosen the lowest-loss elements available in this material system: directional couplers \cite{mikkelsen2014dimensional} for the beamsplitters and thermo-optic phase modulators \cite{2014.opex.harris-galland.heater}.

Figs. \ref{fig:QPP}(c,d) illustrate the primary causes of non-idealities in these devices: in directional couplers, small variations in the dimensions and spacing of coupled waveguides (Fig. \ref{fig:QPP}(c)) result in varied splitting ratios, while in phase shifters, free carrier absorption in the doped silicon regions (Fig. \ref{fig:QPP}(d)) results in increased propagation loss. Our model accounts for realistic variations by using wafer-scale test results for directional couplers \cite{mikkelsen2014dimensional} and phase shifters \cite{2014.opex.harris-galland.heater}. Wafer-scale test data --- as opposed to single-device test data --- improves the validity of our model. We model the splitting ratios by a Gaussian distribution with a mean (standard deviation) of 50\% (4.3\%) \cite{mikkelsen2014dimensional}. We assume the loss in each thermo-optic modulator is also sampled from a non-negative Gaussian distribution \footnote{The continuum limit of a Poisson distribution of scattering events with large mean} with a mean (standard deviation) of 5.16\% (2.84\%). While we vary only two phase shifters in each MZI, we include four phase shifters in the design to balance loss (see Fig. \ref{fig:QPP}(b)).

To incorporate these errors into simulations of QPP performance, we need to modify $U_{\text{MZI}}$.
First, to account for unbalanced splitting ratios, we make the replacement
\begin{equation*}
\frac{1}{\sqrt{2}}
\left(
\begin{matrix}
1 & i\\
i & 1
\end{matrix}
\right)
\rightarrow
\left(
\begin{matrix}
\sqrt{t} & i\sqrt{1-t}\\
i\sqrt{1-t} & \sqrt{t}
\end{matrix}
\right)
\end{equation*}
for each directional coupler, where the value of the transitivity $t$ is chosen randomly according to the distribution above.

To incorporate losses --- if we wish the analysis to remain unitary --- it is necessary to add an additional mode for each lossy component. Then, loss is simply introduced as a beamsplitter with reflectivity equal to the loss. However, due to the block structure of the resulting matrix along with the post-selected nature of the quantum gates we simulate, we can instead work only with the $2 \times 2$ sub-matrix corresponding to the waveguide modes. As such, each diagonal element in the $2 \times 2$ phase-shift matrices acquires a factor of $\sqrt{1-\gamma}$, where the values of the $\gamma$s are distributed according to the loss distribution given above:
\begin{equation*}
\left(
\begin{matrix}
e^{i \phi} & 0\\
0 & 1
\end{matrix}
\right)
\rightarrow
\left(
\begin{matrix}
\sqrt{1-\gamma_1} e^{i \phi} & 0\\
0 & \sqrt{1-\gamma_2}
\end{matrix}
\right)
\end{equation*}

\section{High Fidelity Quantum Gates on the QPP}
\subsection{Individual Quantum Gates}
To demonstrate linear optical quantum gates in the QPP architecture, Figs. \ref{fig:QPP}(e,f) show the post-selected linear optical CNOT gate previously implemented in a custom, static PIC \cite{2003.obrien.branning.cnot,2008.Science.OBrien.quantum_circuit} and the same gate programmed into a subset of the (ideal) QPP lattice, respectively. The beamsplitting ratio of each MZI in Fig. \ref{fig:QPP}(f) is given by $\eta \equiv \sin^2(\theta)$.

This gate, as well as those discussed later, uses the well known dual-rail encoding, i.e. each qubit is encoded in the photon amplitudes in a pair of modes \cite{KLM01}. The control (target) modes are labeled $c_0$ and $c_1$ ($t_0$ and $t_1$) in the figure. The gate succeeds if and only if a single photon is detected in each pair of modes.
Experimental realizations of this gate have demonstrated the promise of PICs, but imperfections in fabrication likely contributed to the reduction in gate fidelities (e.g., to $ 94$\% in Ref. \cite{2008.Science.OBrien.quantum_circuit}). To analyze realistic performance in a QPP system, we simulated 1000 QPPs with splitting ratios and losses given by the aforementioned fabrication model. We then programmed the CNOT gate into each QPP.

To evaluate the performance of each gate, each simulation calculates the 
sub-matrix corresponding to the input and output computational modes:
\begin{align*}
\frac{1}{2}
\left(
\begin{matrix}
e^{i \phi}\sqrt{(1-\gamma_3)t_2} & i e^{i \phi}\sqrt{(1-\gamma_3)(1-t_2)}\\
i\sqrt{(1-\gamma_4)(1-t_2)} & \sqrt{(1-\gamma_4)t_2}
\end{matrix}
\right)\\
\times\left(
\begin{matrix}
e^{i \theta}\sqrt{(1-\gamma_1)t_1} & i e^{i \theta}\sqrt{(1-\gamma_1)(1-t_1)}\\
i\sqrt{(1-\gamma_2)(1-t_1)} & \sqrt{(1-\gamma_2)t_1}
\end{matrix}
\right).
\end{align*}
In simulations, the splitting ratios and losses are determined from a Monte Carlo process; these values can be experimentally determined for a real system using methods presented in \cite{seeSI}. This sub-matrix can then be used to calculate the $4 \times 4$ transform in the computational (i.e. two-qubit) basis $\left\{\ket{\mathbf{00}},~\ket{\mathbf{01}},~\ket{\mathbf{10}},~\ket{\mathbf{11}} \right\}$ \cite{2011.ACM.Aaronson.boson_sampling}, after post-selection. This is then compared to the ideal transformation with the Hilbert-Schmidt inner product \cite{nielsen-chuang} $F(V, V_0)=|V^\dagger V_0|^2$, where $V_0$ is the ideal $4 \times 4$ transform and $V$ is the calculated transform. Normalization (corresponding to post-selection) is performed by scaling $V$ and $V_0$ such that $F(V,V)=F(V_0,V_0)=1$.

\begin{figure}[tb]
\includegraphics[width=3.3in]{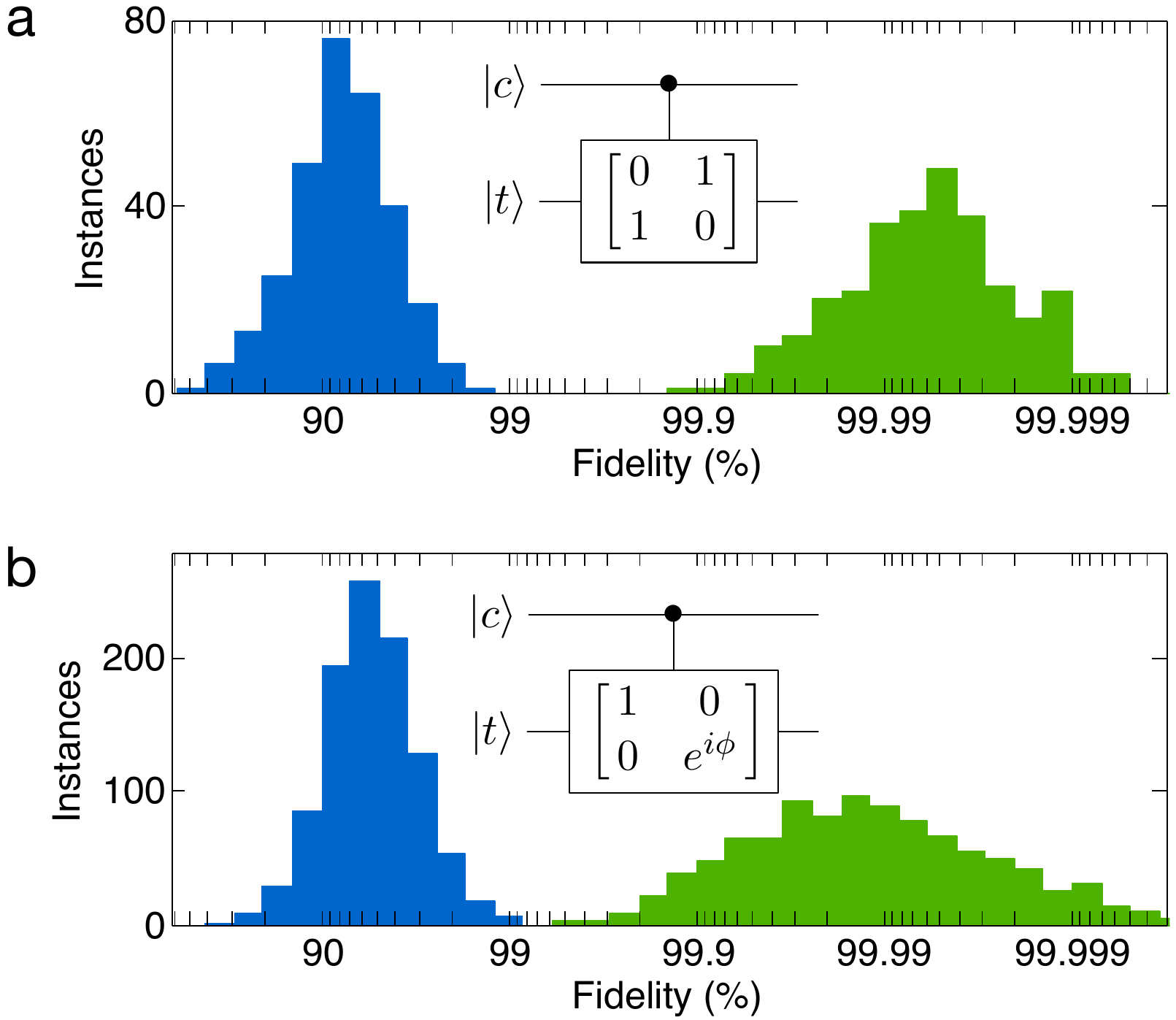}
\caption{{\footnotesize
(a) Performance of the CNOT gate for 1000 instances of the QPP. The blue (green) histogram plots the fidelity before (after) optimization of the phase settings.
(b) Results pre- and post-optimization for the CPHASE gate over 300 instances of the QPP. For each simulation, the reported fidelity is the minimum over six different choices of $\phi$ (the phase applied by the controlled operation), equally distributed from 0 to $2\pi$.}}

\label{fig:CNOT}
\end{figure}

The blue histogram in Fig. \ref{fig:CNOT}(a) shows the fidelity of the CNOT gate over the 1000 simulated QPPs, without optimal MZI tuning.
These simulations yield a median fidelity of 94.52\%, which is similar to experimentally reported values in custom PICs (e.g., \cite{2011.Thompson.OBrien.review, 2008.Science.OBrien.quantum_circuit}). We then performed a nonlinear optimization \cite{RinnooyKan_Timmer_nlopt, Kucherenko_Sytsko_nlopt,2014.Forthcoming.Steinbrecher.Optimization} of the MZI phase settings \cite{QPPnote_reck} to maximize this fidelity for each instance of disorder \cite{seeSI}. The green histogram in Fig. \ref{fig:CNOT}(a) shows the optimized QPP performance; the median fidelity improved dramatically to $99.99\%$.

We performed the same tuning procedure on the post-selected CPHASE gate of Ref. \cite{kieling2010photonic}, for which we observe a similar improvement in median fidelity after optimization from 92.22\% to 99.99\% (see Fig. \ref{fig:CNOT}(b)). These results show that post-fabrication optimization enables the reliable implementation of high-fidelity quantum logic gates on QPPs using currently realizable PICs.

\subsection{Iterative Phase Estimation Algorithm}
The possibility of high-fidelity operations makes the QPP architecture attractive for studying larger-scale quantum algorithms. As it is dynamically reconfigurable, it is well suited for iterative algorithms that rapidly update the circuit in response to previous measurements. Here, we examine the performance of one such algorithm, the iterative phase estimation algorithm. The IPEA is an iterative procedure used to solve for the eigenvalues of a Hamiltonian, which has applications in sensing and simulation, which has applications in sensing and simulation \cite{2012.NPhys.Walther.photonic_quantum_sim}. The IPEA maps a Hamiltonian $H$ to a unitary propagator $U \equiv e^{iH\tau}$. In this approach, solving the eigenvalue problem $U\ket{u}=e^{i2\pi\lambda}\ket{u}$ is equivalent to calculating the energy levels of $H$. A binary expansion of $\lambda$ can be calculated by adaptive and iterative bitwise measurements \cite{2005.Science.Aspuru-Guzik.qu_sim_molecular_energy,2011.Whitfield.Aspuru-Guzik.IPEA}.

\begin{figure}[tb]
\includegraphics[width=3.2in]{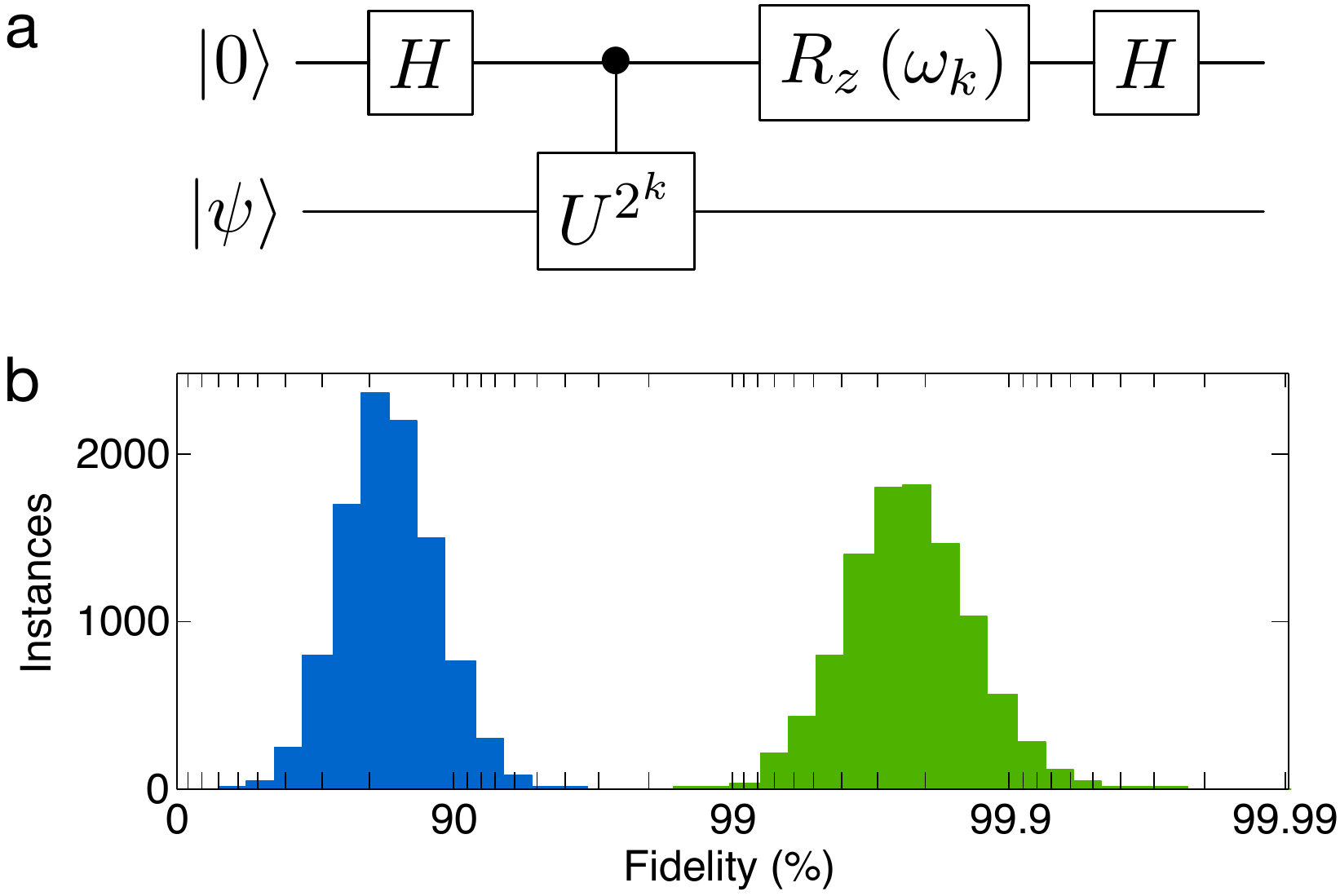}
\caption{{\footnotesize  (a) Quantum circuit for the IPEA, as outlined in Ref. \cite{Lanyon:2010jf}. (b) The IPEA fidelity with unoptimized (blue) and optimized (green) performance. By optimizing the circuit to account for fabrication imperfections, the QPP enables very high process fidelities. Again, note the logarithmic scaling to capture both unoptimized and optimized performance on the same axes.}}
\label{fig:IPEA}
\end{figure}

Fig. \ref{fig:IPEA}(a) shows the quantum circuit for the two qubit IPEA; as demonstrated in \cite{Lanyon:2010jf}, this is sufficient to calculate the first four energy levels of an H$_2$ molecule over a range of atomic separations. This is achieved through the use of a basis set in which the Hamiltonian is block-diagonal with at most $2 \times 2$ blocks. To simulate the performance of this system on the QPP, we decompose the controlled unitary of the IPEA into a CPHASE gate with additional single-qubit rotations. We then split the system into three sections that were optimized separately: the input single-qubit rotations, the CPHASE gate, and the output single-qubit rotations. This decomposition into individually optimized gates is useful for computational efficiency.

We find that for 10,000 simulated instances of the QPP, the unoptimized IPEA performed with a median fidelity of $82.63\%$ (Fig. \ref{fig:IPEA}(b), blue). When using our optimized gates, the median fidelity rose to $99.77\%$ (Fig. \ref{fig:IPEA}(b), green). While only two qubits are required for small simulations, such as an H$_2$ molecule, more qubits are required for larger systems, motivating the development of large-scale PICs such as the QPP.

\section{Quantum Random Walks}

\begin{figure*}[tb]
\includegraphics[width=6.8in]{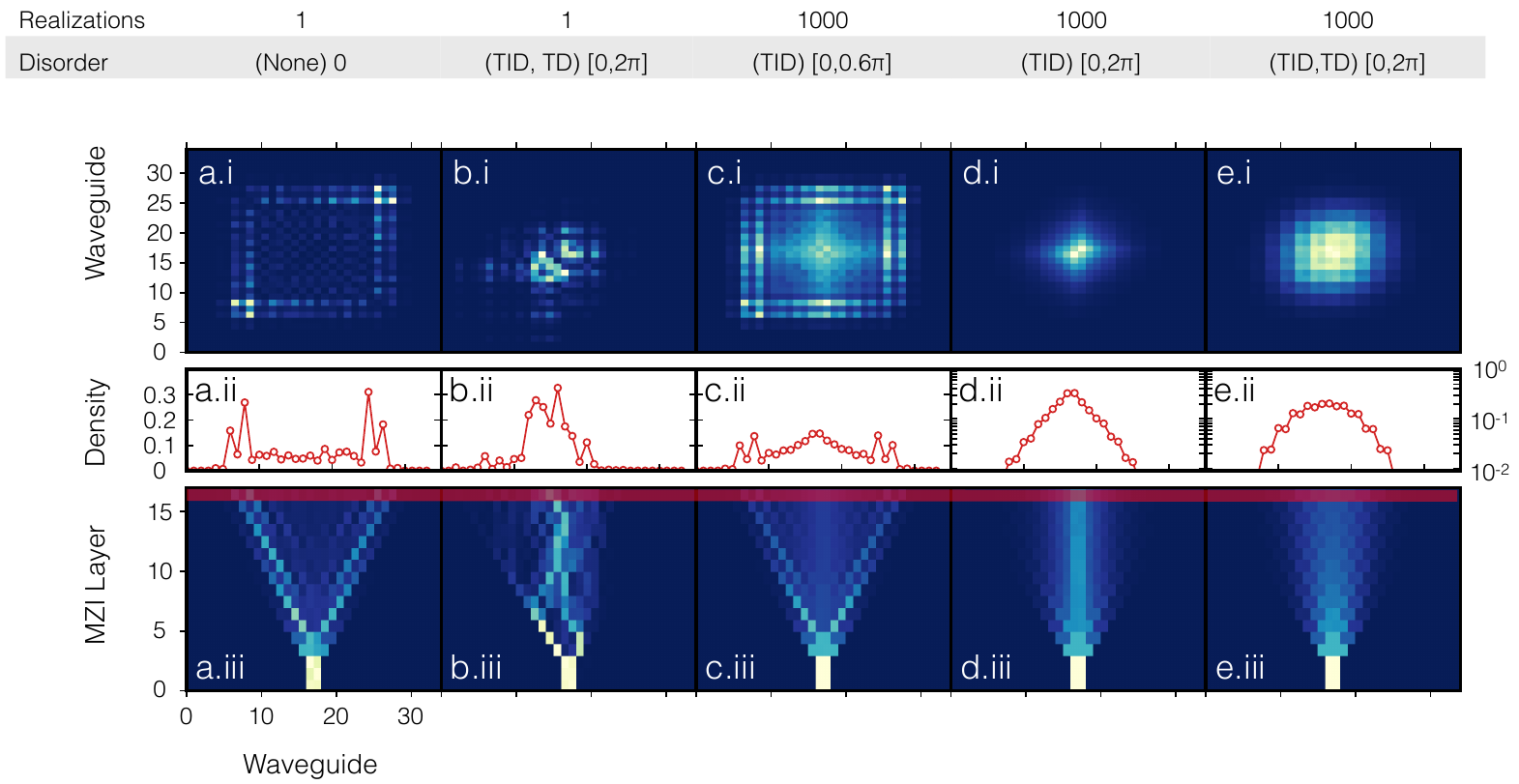}
\caption{{\footnotesize
A simulation of the DTQRW in the QPP post-selected on detecting two photons for various levels of time-dependent (TD) and time-independent (TID) disorder.
(a.i-e.i) Correlation functions for output waveguide positions in the QPP lattice.
(a.ii-e.ii) Particle density distributions as a function of waveguide position (same as the last layer of a.iii-e.iii, marked in red). d.ii and e.ii have logarithmic scales.
(a.iii-e.iii) Dynamics of QRW where the $x$-axis and $y$-axis represent the waveguide output position and MZI layer, respectively.
(a.i-iii) Propagation of input state $(\ket{20}_\text{LR} + \ket{02}_\text{LR})/\sqrt2$ revealing bunching effect seen for continuous-time QRWs.
(b.i-iii) A single realization of TID and TD disorder in the QPP resulting in highly irregular propagation.
(c.i-iii) Average of 1000 realizations of weak TID disorder showing the coexistence of bunching and localization.
(d.i-iii) Average of 1000 realizations of TID disorder showing an exponential density distribution --- the hallmark of Anderson Localization.
(e.i-iii) Average of 1000 realizations of TID and TD disorder, showing delocalization and a Gaussian distribution.
}}
\label{fig:QRW}
\end{figure*}

The programmability of the QPP also enables detailed studies of single- \cite{PhysRev.109.1492,segev.2007.ander,PhysRevLett.100.013906} and multi-photon \cite{PhysRevLett.102.253904,2010.Peruzzo.OBrien.QRW, PhysRevLett.105.163905,ander.PIC.2013,2013.Giuseppe.Saleh.Anderson} quantum random walks on a lattice with discrete, nearest-neighbor coupling. QRWs are attractive for their application to the problems of quantum simulation \cite{2008.JCP.Mohseni.qu_sim_photosynth}, database search \cite{Childs_Goldstone_2004}, and Boson Sampling \cite{2011.ACM.Aaronson.boson_sampling}.

In the discrete-time QRW, a particle with an internal binary degree of freedom (a ``coin")  is placed on the lattice. At each step of the walk, two operations occur: the internal state of the coin is prepared and the particle is shifted left or right (as indicated in Fig. \ref{fig:QRW}) according to the state of the coin (``left" and/or ``right"). We use a spatial encoding for both the position and the coin state of a quantum walker: position is defined at an MZI, while the coin state is defined by occupation between the two output waveguides of the MZI. The coin toss operation is controlled by the MZI splitting ratio and output phase. The MZI lattice implements a shift operation where photons in the left (right) state propagate left (right) to the next layer of the QPP. The rectangular lattice shown in Fig. 1(a) can implement such a QRW more compactly than the triangular lattice proposed in Ref. \cite{1994.PRL.Reck-Zeilinger-Bartani.any_unitary_operator}.

We studied the propagation of two indistinguishable photons on a QRW in the QPP. The path-entangled initial state is $\ket{\psi}_i=(\ket{20}_\text{LR} + \ket{02}_\text{LR})/\sqrt{2}$, where \text{L} and \text{R} are the two outputs of the first MZI of the QRW, MZI$_1$. This state is prepared in the QPP by first launching indistinguishable photons into the two waveguides of the first MZI set to $(\eta,\phi)=(0.5,\pi/2)$, producing the desired state $\ket{\psi}_i$. The next layer of MZIs is set to $(\eta,\phi)=(1,0)$ and $(1,0)$ in order to route the state to the first layer of the random walk. The state is then evolved in the following 15 MZI layers of the QPP, where all internal phases $\theta$ are set to $\pi/2$. In these simulations, disorder is introduced by sampling the MZI output phases ($\phi$) randomly from a uniform distribution on the interval $[0,\Phi_\text{max}]$.

We first consider a lattice without disorder, i.e., $\Phi_\text{max} = 0$. Simulation results for a realistic QPP are plotted in Figs. \ref{fig:QRW}(a.i-iii). Fig. \ref{fig:QRW}(a.i) shows the two-photon correlation function, (a.ii) plots the particle density at the output, and (a.iii) shows the particle density at every layer of the QPP. The two-photon correlation function (Fig. \ref{fig:QRW}(a.i)) displays stronger correlations for neighboring waveguides (``bunching") and particle density peaks at the edges of the array (a.ii,iii). This bunching phenomenon is analogous to Hong-Ou-Mandel interference observed for two input and two output modes \cite{1987.PRL.Hong-Ou-Mandel.interference}. An analogous effect is seen in continuous-time QRWs for two indistinguishable photons launched in neighboring waveguides \cite{PhysRevLett.102.253904,2010.Peruzzo.OBrien.QRW}.

The impact of disorder on path-entanglement and the transport of multi-photon states is not presently well understood, and remains an active area of research. A single realization of disorder offers little information as it can contain extreme arrangements not representative of the majority of lattices with the same level of disorder. This can be seen by comparing a single realization of disorder (Fig. \ref{fig:QRW} (b.i-iii)) to 1000 realizations of disorder (e.i-iii), for $\Phi_\text{max}=2\pi$ in both cases. To build robust statistics, multiple instances of a given level of disorder are required. Until now, this could have been accomplished by fabricating numerous samples or by post-processing PICs \cite{segev.2007.ander,PhysRevLett.100.013906,ander.PIC.2013,2013.Giuseppe.Saleh.Anderson}. This approach is difficult to extend to hundreds or thousands of instances. While fast switches could be used to modulate photons passing through a looped QRW \cite{2011.Schreiber.Silberhorn.DisorderQRW}, there are significant losses associated with this setup that hinder its application to large-scale experiments.

However, a single QPP could generate many instances of disorder. Time-dependent (independent) disorder can be realized with random phase settings along (orthogonal to) the direction of propagation. Applying weak time-independent disorder ($\Phi_\text{max}=0.6\pi$) to the lattice results in two-photon correlation and density functions that exhibit both bunched and localized characteristics (Fig. \ref{fig:QRW}(c.i-iii)). This effect was predicted for continuous-time QRWs \cite{PhysRevLett.105.163905}.

Strong, time-independent disorder in the QPP lattice ($\Phi_\text{max}=2\pi$) reveals the characteristic exponential distributions of Anderson localization (Fig. \ref{fig:QRW}(d.i-iii)). The incorporation of time-dependent disorder results in the two-photon correlation function and particle density distribution transitioning from exponential localization to Gaussian delocalization (Fig. \ref{fig:QRW}(e.i-iii)) --- indicative of a crossover to diffusion \cite{Amir:2009uo,2012.Levi.Segev.HyperTransport}. Although fabrication defects were included in the simulations, we find that the two-photon correlations and densities were largely unaffected \cite{seeSI}.

\section{Discussion}

We have shown that a QPP, fabricated in current silicon photonics processes, could enable high-fidelity quantum gates and quantum simulation. We focused on post-selected gates to compare this system to preceding PIC-based experiments. Looking forward, one goal of linear optical quantum computing (LOQC) systems is to achieve the error threshold necessary for fault-tolerant quantum computation \cite{KLM01}. For post-selected LOQC, this threshold can be as high as ~1\% \cite{knill2005}, but with limitations on overhead (e.g., $<10^4$ physical CNOT gates per qubit and gate), the error rate must be much lower: $\sim 10^{-3}-10^{-4}$ \cite{knill2005}. The optimization work presented above, in combination with advanced silicon processes, offers a path toward achieving these demanding error rates in the QPP architecture.

Proposed universal quantum computers based on LOQC will also require efficient single-photon sources, single-photon detectors, and feed-forward operations. There has been rapid progress integrating these elements into the silicon photonics platform; recent examples include entangled-photon sources based on four-wave mixing \cite{2010.chen.migdal} and waveguide-integrated superconducting single-photon detectors \cite{2013.Najafi-Mower.PIC-SNSPD,2011.Pernice.Tang.SNSPD}. The potential for multiplexing the emission of spontaneous single-photon sources \cite{PhysRevA.66.053805,2011.PRA.Mower.AMPP} could enable high-efficiency state preparation for quantum computation; low-latency superconducting logic \cite{mccaughan2014superconducting} could enable the feed-forward required for scalable LOQC; and low photon-number nonlinear elements could enable photon-photon interaction and deterministic quantum logic \cite{PhysRevLett.111.247401,Englund2008}.

The high-dimensional transformations possible on the QPP could also enable a number of applications in classical optics, including multi-input multi-output, transparent, non-blocking switches \cite{Yang:11,Chen:2014el}, signal routers, high-dimensional beam splitters, and large phased arrays \cite{sun2013large}, e.g., for LIDAR applications.

\section{Conclusion}

We presented a detailed analysis of the feasibility of a reconfigurable quantum photonic processor that enables high-fidelity linear optical transformations and could greatly accelerate prototyping of quantum algorithms in integrated quantum photonics. As demonstrated by our simulation of quantum walks, reconfigurability also enables a single device to perform statistically robust studies of the propagation of photons through complex optical networks. The predicted high fidelity of quantum operations under realistic fabrication defects suggests that a QPP reaching high post-selected gate fidelities is within experimental reach.

\begin{acknowledgements}
We would like to acknowledge funding from the AFOSR MURI program under grant number (FA9550-14-1-0052).
J.M. acknowledges support from the iQuISE fellowship.
N.H. acknowledges that this material is based upon work supported by the National Science Foundation Graduate Research Fellowship under Grant No. 1122374.
G.R.S. was supported by the Department of Defense (DoD) through the National Defense Science \& Engineering Graduate Fellowship (NDSEG) Program. D.E. acknowledges support from the Sloan Research Fellowship in Physics.
Y.L. acknowledges support from the Pappalardo Fellowship in Physics.
\end{acknowledgements}

\bibliographystyle{apsrev_no_links.bst}
\bibliography{references_BibDesk7_Dirk2}

\begin{thebibliography}{60}
\bibitem[{QPP()}]{QPPnote_reck}
\bibinfo{howpublished}{Neither the optimized nor unoptimized CNOT gate
  programmed into the QPP directly implemented the decomposition of the
  corresponding single-photon unitary transformation proposed in Ref. [18]. The
  decomposition in Ref. [18] was not shown to be optimal.}
\end{thebibliography}

\clearpage


\section*{Supplementary material}

# QUANTUM RANDOM WALKS: STATE PREPARATION

The unit cell of the QPP (the MZI) does not implement a symmetric beamsplitter and therefore realizes an asymmetric quantum walk. In this section, we will consider one possible method for realizing a symmetric quantum walk. With MZI phases set to $\theta = \pi/2$ and $\phi = 0$, the following unitary (Hadamard) transformation is applied to the input modes (to a global phase),

$$U = \frac{1}{\sqrt{2}} \begin{pmatrix} 1 & 1 \\ 1 & -1 \end{pmatrix}.$$

Thus, photons incident from the left port acquire a different phase compared to those incident from the right port (a phase difference of $\pi$). It is possible to correct for the asymmetric action of the MZI beamsplitter during the quantum walk by injecting a state in an equal superposition of the input modes; namely a NOON state with N = 2.

Given two indistinguishable photons, this state can be prepared by the QPP. One of the photons is launched into port R of the blue MZI and the other into port L of the red MZI (Figs. 1(a,b)) — both of which are configured to implement "wires" with a variable output phase shift ($\theta = \pi$ while varying $\phi$). To generate the NOON state described above, the red MZI is applies a relative phase shift of $\phi = \pi/2$ while the blue MZI is set to apply $\phi = 0$ relative phase shift. All other MZIs, marked in gray in Figs. 1(a,b), are set to apply the Hadamard operation with $\theta = \pi/2$ and $\phi = 0$. Thus, after the second layer, the state $(|20\rangle_{LR} + |02\rangle_{LR})/\sqrt{2}$ is prepared, where L and R are the two outputs of an MZI (Fig. 1(c)).

In Fig. 2, we reconsider the quantum walk simulations from the main text for the case of ideal beamsplitters and lossless phase shifters. In all cases, the differences between the non-ideal distributions and correlations and the ideal ones are minimal.

# CHARACTERIZING A QPP

The optimization algorithm presented in the main text was given access to the 2 × 2 matrix for each MZI as a function of $\theta$ and $\phi$. Here, we discuss ways an experimentalist could characterize a QPP system to extract sufficient information about each MZI to inform these optimizations. To start, we present a method that is algorithmically simple but involves augmenting the QPP unit cell to include detectors that can be "switched off," for example by using tunable ring resonators to create a switched drop filter. These would be placed at the outputs of each MZI (e.g. the outputs at the top of Fig. 1(a)). Such additions would add considerable fabrication

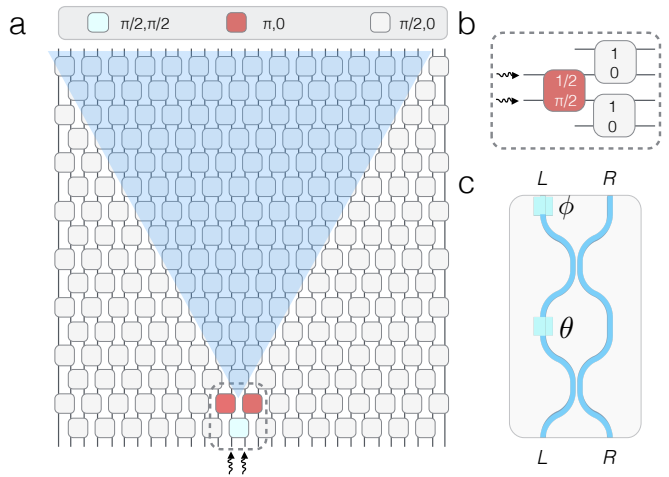

FIG. 1. (a) Sixteen layer QPP model used for the simulation of ballistic propagation in an MZI Lattice. The green triangular region shows the light cone for the quantum walk after state preparation. (b) The first two layers of the QPP are used to prepare the input state (image rotated 90 degrees clockwise with respect to dashed region in (a)). Indistinguishable photons are injected into port R of the blue MZI and port L of the red MZI. These MZIs are configured to act as straight waveguides. The output phase shifter, $\phi$ for the red (blue) MZI is set to $\pi/2$ (0). As before, the top number in the MZI box represents the splitting ratio $\eta$ and the bottom number represents the output phase $\phi$. (c) Schematic view of MZI showing phase setting labels.

and systematic overhead to a QPP — to provide an alternative, we discuss a more involved method at the end of the section that uses only detectors at the outputs of the array.

In the notation of Sections II and III, the 2 × 2 transform of a given MZI is

$$V = \frac{1}{2} \begin{pmatrix} e^{i\phi}\sqrt{(1-\gamma_3)t_2} & ie^{i\phi}\sqrt{(1-\gamma_3)(1-t_2)} \\ i\sqrt{(1-\gamma_4)(1-t_2)} & \sqrt{(1-\gamma_4)t_2} \end{pmatrix}$$
$$\times \begin{pmatrix} e^{i\theta}\sqrt{(1-\gamma_1)t_1} & ie^{i\theta}\sqrt{(1-\gamma_1)(1-t_1)} \\ i\sqrt{(1-\gamma_2)(1-t_1)} & \sqrt{(1-\gamma_2)t_1} \end{pmatrix},$$

where $\gamma_1$ and $\gamma_3$ correspond to the losses in the $\theta$ and $\phi$ modulators, respectively, and $\gamma_2$ and $\gamma_4$ correspond to the other two (loss balancing, but otherwise inactive) modulators. We can capture $V$ compactly as an arbitrary 2 × 2 complex matrix:

$$V = \begin{pmatrix} a & b \\ c & d \end{pmatrix} = \begin{pmatrix} a_r e^{ia_p} & b_r e^{ib_p} \\ c_r e^{ic_p} & d_r e^{id_p} \end{pmatrix},$$

where $x_r = |x|$ and $x_p = \arg(x)$ for $x \in \{a, b, c, d\}$, the $x_r$s are functions of $\theta$ and the $x_p$s are functions of both $\theta$ and $\phi$. Then, our problem reduces to determining these eight real parameters as a function of the phases $\theta$ and $\phi$.

We can extract the $x_r$ elements directly using the embedded detectors; assuming that only one of the input op-

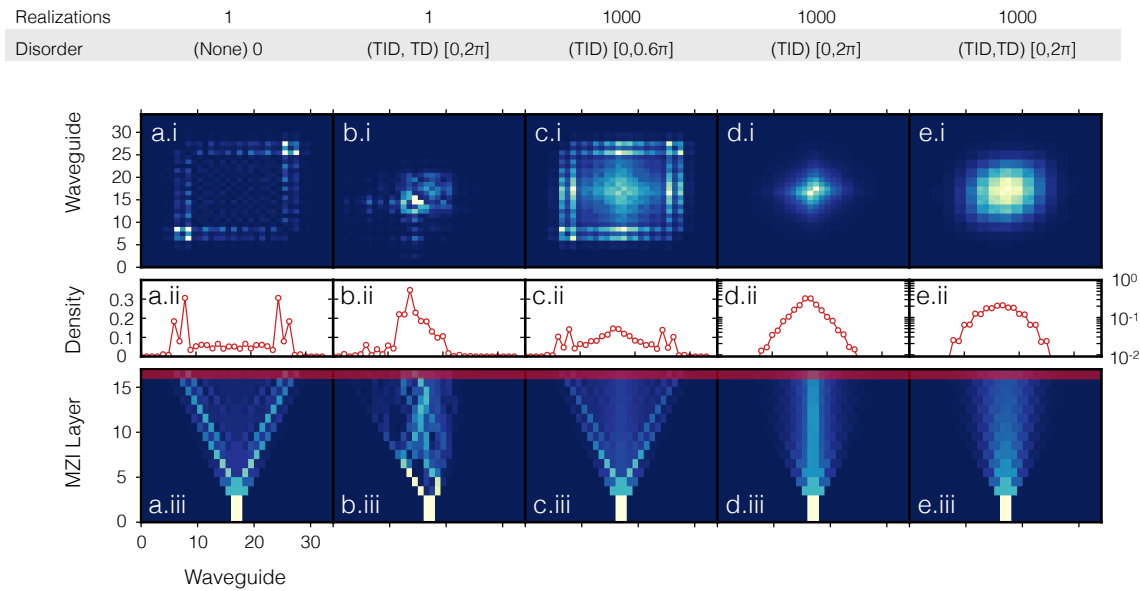

FIG. 2. A simulation of the discrete-time quantum walk in an *ideal* QPP, for various levels of time-dependent (TD) and time-independent (TID) disorder. (a.i-e.i) Two-particle correlation functions for output waveguide positions in the QPP lattice. (a.ii-e.ii) Particle density distributions as a function of waveguide position (same as the last layer of a.iii-e.iii, marked in red). (a.iii-e.iii) Dynamics of a two photon QRW where the $x$-axis and $y$-axis represent the waveguide output position and MZI layer, respectively. (a.i-iii) Propagation of input state $(|20\rangle_{LR} - |02\rangle_{LR})/\sqrt{2}$ revealing bunching effect seen for continuous-time QRWs. (b.i-iii) A single realization of TID and TD disorder in the QPP resulting in highly irregular propagation. (c.i-iii) Average of 1000 realizations of weak TID disorder showing the coexistence of bunching and localization. (d.i-iii) Average of 1000 realizations of TID disorder showing an exponential distribution — the hallmark of Anderson Localization. (e.i-iii) Average of 1000 realizations of TID and TD disorder, showing delocalization and a Gaussian distribution.

tical powers $P_{\text{in,top}}$ and $P_{\text{in,bottom}}$ is non-zero for a given measurement,

$$a_r(\theta) = \sqrt{P_{\text{out,top}}(\theta)/P_{\text{in,top}}} \quad (1)$$
$$b_r(\theta) = \sqrt{P_{\text{out,top}}(\theta)/P_{\text{in,bottom}}} \quad (2)$$
$$c_r(\theta) = \sqrt{P_{\text{out,bottom}}(\theta)/P_{\text{in,top}}} \quad (3)$$
$$d_r(\theta) = \sqrt{P_{\text{out,bottom}}(\theta)/P_{\text{in,bottom}}}. \quad (4)$$

Characterization of the array proceeds iteratively: a known optical power is inserted into each port of the array, and the corresponding matrix values are measured as a function of the relevant $\theta$. This then lets us prepare a known optical power at the inputs to the second layer, which, once characterized, allows for known optical powers at the third, etc. until we have characterized the entire array.

This leaves the determination of the $x_p$ parameters. Using the previous results, we can route light in "wire-paths" throughout the QPP array, where the light travels along a single path from input to output. Externally, the light from this path can then be interfered on a beam-splitter with a local oscillator, giving a phase. In a wire, each MZI is either in the "identity" state or in the "swap" state, meaning there are 8 $x_p$ free parameters per MZI. The total phase acquired along a wire-path is a simple sum of the $x_p$ elements along that path meaning that, so long as there are more wire-paths than $x_p$ values, we can determine all of them by linear regression. In fact, in a given QPP, there are far more ways of constructing a "wire" through the array than there are free parameters (i.e. we have an overcomplete set of equations). For example, the QPP of Figure 1 in the main text has 60 MZIs (giving 480 $x_p$ values) while there are 2976 wire paths from inputs to outputs.

However, such wire-paths are not true wires. Due to imperfections, there will be small amounts of light that travel along other paths yet still reach the target output port. We isolate this light and remove it from our calculations by varying the voltage applied to all modulators in the array not along the wire-path so that this spurious light appears in the Fourier transform of the output signal at a non-zero frequency. This effectively "tags" the confounding light, allowing it to be removed from the result.

Once the $x_p$ values have been found for the wire-paths, individual modulators can be varied to verify the change of the $x_p$s for intermediate settings of the modulators. Interior ($\theta$) modulators will affect which wire-path the light takes, but as the other modulators are held constant, this does not increase the complexity of the characterization.

Until now, this process has assumed switched detectors embedded into the array, which would increase the de-

mands on the fabrication process and likely introduce extra complication. However, we can remove this necessity in return for some added computational cost and characterization time. In particular, we can determine the phase settings necessary to create wire-like paths without any measurement of intermediate optical powers.

If light is input to only a single port of a QPP array, it can only reach a finite number of output ports in either lateral direction. If we consider this light-cone of reachable ports for a given input MZI, the top-most output mode can only be reached by light leaving the top output port of the first MZI; likewise, the bottom-most output mode can only be reached by light leaving the bottom output port of the first MZI. By only putting light into one mode of the first MZI and monitoring the power at one of these edge modes, we can force the MZI to the "identity" or "swap" configurations. This process can then proceed iteratively through the array, setting each MZI on the path to the appropriate configuration. The logarithms of the magnitudes of the $x_r$ elements along a given path add, meaning we can perform a similar linear regression as discussed for the phases above, characterizing the entire array without the need for embedded detectors. The modulation scheme to remove spurious light from calculations is necessary here as well.

While thousands of measurements for the characterization may seem like a daunting experimental task, any QPP realized in practice would be computer controlled, meaning this process would be entirely automated. And, at the speeds of thermo-optic modulators (>100kHz), the characterization would likely take little time on any given chip.

## NONLINEAR OPTIMIZATIONS

As demonstrated in the main paper, quantum operations on the QPP architecture are sensitive to fabrication defects. Even for single-qubit gates, the induced disorder quickly decreases the fidelity below acceptable limits. However, as demonstrated in detail in [1], it is possible to apply numerical optimization techniques to adjust the applied phases to these devices post-fabrication in an efficient manner, achieving extremely high fidelity operation of single gates. Moreover, even though the optimization is performed only locally for each gate, these improvements in fidelity are maintained at the global scale when cascading operations.

For this work, four different individual networks were optimized: the postselected CNOT and CPHASE gates, as well as the single-qubit rotations necessary at the input and output of the iterative phase estimation algorithm. The optimization process uses the computational basis transform applied by each MZI (a $4 \times 4$ complex, two-photon matrix, $\varphi(U)$ [2], that is a principal submatrix of the full unitary transform) and calculates the fitness of a given phase setting using the Hilbert-Schmidt norm. The optimization process is seeded with the set of phases for an ideal sub-block of the QPP and uses a running time-bounded combination of global and local optimization procedures. In order to improve the fidelity achieved using this procedure, buffer layers of MZIs were added to the input and output of each gate, expanding the size of the network slightly.

The calculation of the computational basis transform is performed as follows. First, a vector corresponding to the phase of each modulator is selected by the optimization algorithm. These phases are used to generate the single particle unitary [2] transform generated by the QPP sub-block under consideration, incorporating fabrication errors. This is then used to calculate the matrix elements of the computational basis transform.



\end{document}